\documentclass[aps,prb,twocolumn,floatfix,showpacs,showkeys]{revtex4}
\usepackage{amsmath}
\usepackage{amssymb}
\usepackage{amsfonts}

\begin{document}
\title{Oscillatory spin relaxation rates in quantum dots}
\author{C. F. Destefani}
\author{Sergio E. Ulloa}
\affiliation{Department of Physics and Astronomy and Nanoscale and
Quantum Phenomena Institute, Ohio University, Athens, Ohio
45701-2979}
\date{\today}

\begin{abstract}
Phonon-induced spin relaxation rates in quantum dots are studied
as function of in-plane and perpendicular magnetic fields,
temperature and electric field, for different dot sizes. We
consider Rashba and Dresselhaus spin-orbit mixing in different
materials, and show how Zeeman sublevels can relax via
piezoelectric and deformation potential coupling to acoustic
phonons. We find that strong lateral and vertical confinements may
induce minima in the rates at particular values of the magnetic
field, where spin relaxation times can reach even seconds. We
obtain good agreement with experimental findings in GaAs quantum
dots.
\end{abstract}

\maketitle

Since the proposal of a qubit based on the electron spin of
quantum dots (QDs),\cite{vicenzo} much work has been done to
understand the processes that may cause their relaxation, since
long coherence times are required. One of those processes is
related to the phonon-induced spin-flip rates of Zeeman sublevels
in QDs in magnetic fields, where the spin purity of the levels is
broken by the spin-orbit (SO) interaction. A recent
experiment\cite{hanson} has shown that spin relaxation time has a
lower bound of 50 $\mu$s at an in-plane field of 7.5 T in a GaAs
QD defined in a 2DEG, while theoretical work\cite{vitaly} has
given 5 $\mu$s at the same QD parameters. In general, SO effects
have been considered via perturbation theory,\cite{nazarov,bulaev}
although exact treatments have also been
presented.\cite{rogerio,cheng,des1} The perturbative approach,
which includes only a few states, has been called into question by
the demonstration that a larger basis is needed in order to
achieve convergence even for the lowest QD states when the QD
vertical width is narrow,\cite{cheng} where a complex interplay
between different energy scales is present.\cite{ZeemanSO} The
influence of high SO coupling on the QD effective $g$-factor has
been recently addressed.\cite{gfactor}

Using exact diagonalization, our goal is to provide a bridge
between the different approaches used so far, so that their range
of validity can be well established in terms of materials
(wide/narrow-gap), QD sizes, and directions of the magnetic field.
For GaAs we show that the perturbative approach is indeed
satisfactory for in-plane fields \textit{for any QD
confinement}, where good agreement is found with calculation\cite%
{vitaly,rogerio} and experiment,\cite{hanson} and where the
\textit{piezoelectric} potential is shown to dominate. If the
field is perpendicular, however, we find that the perturbative
approach\cite{nazarov} indeed fails, giving relaxation rates
orders of magnitude longer than our results \textit{if the
$z$-confinement is strong}.\cite{cheng} For higher perpendicular
fields, we show that \textit{minima} are found in the rates
depending on the QD features. For InSb the \textit{oscillatory
rates} are found for both in-plane and perpendicular fields and
strong confinement. We find that the \textit{deformation}
potential dominates the rates, so that previous
results\cite{rogerio} under in-plane fields not including such
potential yielded rates orders of magnitude smaller than ours. We
show that the oscillatory behavior is caused by the nature of the
$z$-confinement, as well as by the complex interplay between the
physical scales playing a role in the system: Zeeman and SO
energies, as well as QD lateral and vertical sizes.

The QD is defined by an in-plane parabolic confinement, $V(\rho
)=m\omega _{0}^{2}\rho ^{2}/2$, where $m$ ($\omega
_{0}=E_{0}/\hbar $) is the electronic effective mass (confinement
frequency); the QD lateral length is $l_{0}=\sqrt{\hbar /(m\omega
_{0})}$. The vertical confinement $V(z)$ is strong enough so that
only the state in the first quantum well subband is relevant, and
its function is $\varphi _{z}(z)=\sqrt{2/z_{0}}\sin \left( \pi
z/z_{0}\right) $ if a hard wall is assumed, $z_{0}$ being the QD
vertical well thickness. In a magnetic field $\mathbf{B}$, the
unperturbed Hamiltonian,
$H_{0}=\hbar^{2}\mathbf{k}^{2}/2m+V(\rho)+H_{Z}$, has the
well-known Fock-Darwin (FD) solution,\cite{jacak} where
$H_{Z}=g_{0}\mu _{B}\mathbf{B}\cdot\mathbf{\sigma }/2$ is the
Zeeman term. We include all SO terms in 2D zincblende QDs, namely
Rashba\cite{ra} and Dresselhaus\cite{dre} interactions. The former
is due to surface inversion asymmetry (SIA) induced by the 2D
confinement, while the latter is caused by the bulk inversion
asymmetry (BIA) present in
zincblende structures. The SIA Hamiltonian is $H_{SIA}=\alpha%
\mathbf{\sigma }\cdot \mathbf{\nabla }V(\rho ,z)\times%
\mathbf{k}$, while the BIA is $H_{BIA}=\gamma \lbrack \sigma%
_{x}k_{x}\left( k_{y}^{2}-k_{z}^{2}\right) +\sigma _{y}k_{y}\left(%
k_{z}^{2}-k_{x}^{2}\right) +\sigma _{z}k_{z}\left(%
k_{x}^{2}-k_{y}^{2}\right) ]$, with coupling constants $\alpha$
and $\gamma$. The $z$-confinement yields the electric field
$dV/dz$ in the SIA Hamiltonian as well as the momentum average
$\left\langle k_{z}^{2}\right\rangle =(\pi /z_{0})^{2}$ in the BIA
terms. The full QD Hamiltonian is then $H=H_{0}+H_{SIA}+H_{BIA}$,
which is diagonalized in a basis containing $110$ FD states.
Details about the derivation of terms in $H$, as well as their
selection rules for (anti)crossings in the QD energy spectrum, are
found elsewhere.\cite{des1,gfactor}

We calculate spin relaxation rates between the two lowest QD
Zeeman sublevels caused by piezoelectric and deformation acoustic
phonons via Fermi's Golden Rule: $\Gamma _{fi}=2\pi /\hbar
\sum_{j,\mathbf{Q}}\left\vert \gamma _{fi}(\mathbf{q})\right\vert
^{2}\left\vert Z(q_{z})\right\vert ^{2}\left\vert
M_{j}(\mathbf{Q})\right\vert ^{2}(n_{Q}+1)\delta \left( \Delta
E+\hbar c_{j}Q\right) $, where the sum is over the emitted phonon
modes $j$ ($j=LA,TA1,TA2$) with momentum
$\mathbf{Q}=(\mathbf{q},q_{z})$. The term $Z(q_{z})=\left\langle
\varphi _{z}\right\vert e^{iq_{z}z}\left\vert \varphi
_{z}\right\rangle $ ($\gamma _{fi}(\mathbf{q} )=\left\langle
f\right\vert e^{i\mathbf{q}.\mathbf{r}}\left\vert i\right\rangle
$) is the form factor perpendicular (parallel) to the 2D-plane
(position is $\mathbf{R}=(\mathbf{r},z)$), while $n_{Q}$ is the
phonon distribution with energy $\hbar c_{j}Q$; energies $ \Delta
E=\varepsilon _{f}-\varepsilon _{i}$ and states $\left\vert
i\right\rangle $, $\left\vert f\right\rangle $ are obtained via
diagonalization of the total $H$, so that the SO mixing is fully
taken into account. The element $M_{j}(\mathbf{Q})=\Lambda
_{j}(\mathbf{Q})+i\Xi _{j}( \mathbf{Q})$ includes both
piezoelectric $\Lambda _{j}$ and deformation $\Xi _{j}$
potentials; in zincblende structures\cite{pho} and in cylindrical
coordinates, they become $\Xi _{LA}(\mathbf{Q})=\Xi
_{0}A_{LA}\sqrt{Q}$ (only $LA$ is present for $\Xi _{j}$),
$\Lambda _{LA}(\mathbf{Q})=3\Lambda _{0}A_{LA}\sin (2\theta
)q^{2}q_{z}/2Q^{7/2}$, $\Lambda _{TA1}(\mathbf{Q} )=\Lambda
_{0}A_{TA}\cos (2\theta )qq_{z}/Q^{5/2}$, and $\Lambda _{TA2}(
\mathbf{Q})=\Lambda _{0}A_{TA}\sin (2\theta )\left(
2q_{z}^{2}/q^{2}-1\right)q^{3}/2Q^{7/2} $ (both $TA1$ and $TA2$
modes are compacted as a single $TA$ mode for $\Lambda _{j}$),
where $A_{j}=\sqrt{\hbar (2N_{0}Vc_{j})^{-1}}$ and $\Lambda
_{0}=4\pi eh_{14}/\kappa $. The bulk phonon constants are $\Xi
_{0}$ and $eh_{14}$, $ c_{j}$ are the sound velocities ($c_{TA1}$
= $c_{TA2}$ = $c_{TA}$ $\neq$ $c_{LA}$), $\kappa $ is the
dielectric constant, and $N_{0}$ is the electron density. The
triple space integration ($[r,\phi_{r},z]$) yields an analytical
solution,\cite{cheng} and we were able to do two
($[\phi_{q},q_{z}]$) out of the momentum integration, leaving a
numerical integral only in $q$. For a later use, the only
$z_{0}$-dependence in this remaining integral in $\Gamma _{fi}$
reads $F_{j}(z_{0})=(d_{j}z_{0}-(d_{j}z_{0})^{3}/\pi^{2})^{-2}
\sin^{2}(d_{j}z_{0})$, where $d_{j}=\sqrt{(\Delta E/\hbar
c_{j})^{2}-q^{2}}/2$; $q$ runs from $0$ to $\Delta E/\hbar c_{j}$,
while $F_{j}(z_{0})$ is multiplied by polynomials and exponentials
in $q$ in the total $\Gamma _{fi}$. No approximation is needed in
our derivation of $\Gamma _{fi}$, so that the 3D nature of the
phonon is taken into account and the full form factor
$e^{i\mathbf{Q}.\mathbf{R}}$ is used.

Panel $A$ of Fig.\ \ref{fi1} shows the phonon-induced spin-flip
rates as a function of in-plane field, $B_{\parallel }$, for
different values of $E_{0}$ (solid lines) for GaAs
QDs.\cite{parga} Larger $E_0$ (smaller QDs) present smaller
spin-flip rates, i.e., longer relaxation times. This happens
because as $E_{0}$ increases, the orbital levels become more
separated and then the SO coupling becomes relatively less
important. A strong dependence of the rates with $B_{\parallel }$
and $E_{0}$ is clear. For example, at $B_{\parallel }=2$ T, one
gets $2.5\times 10^{2}$ s$^{-1}$ at $E_{0}=0.7$ meV and $2.5\times
10^{-1}$ s$^{-1}$ at $E_{0}=5.0$ meV; at $B_{\parallel }=15$ T the
rate goes from $5.2\times 10^{6}$ s$^{-1}$ to $6.9\times 10^{4}$
s$^{-1}$ for those same values of $E_{0}$. Finite temperature
(dashed-dotted line for $T=12$ K) enhances the rates by about one
order of magnitude due to the enhancement of the phonon
population. Larger wells (dotted line for $z_{0}=100$ \AA) have
rates about one order of magnitude smaller because the BIA term
($\left\langle k_{z}^{2}\right\rangle \approx 1/z_{0}^{2}$)
decreases. The same $T$ and $z_{0}$ behavior is found for any
$E_{0}$. Larger electric fields (dashed line for $dV/dz=-38$
meV/\AA) have little effect on the rates for large $E_0$, but at
smaller $E_{0}$ the rates decrease by one order of magnitude; this
means that by increasing the SIA term so that it equals the BIA
coupling, the rates can \textit{decrease} even though the spin
mixing gets stronger. For \textit{GaAs} QDs under
\textit{in-plane} fields, we find that the \textit{TA
piezoelectric} coupling dominates at \textit{low fields}
($\lesssim 14$ T for $E_{0}=5$ meV), while at high fields the
deformation potential takes over (symbols in Fig.\ \ref{fi1}$A$).

\begin{figure}
\caption{$(A)$. Zeeman sublevel spin-flip rates for GaAs
QDs\cite{parga} under in-plane magnetic field for different QD
confinements (solid lines). Dashed-dotted, dotted, and dashed
lines respectively show influence of higher temperature, well
width, and electric field on the $3.5$ meV QD. Symbols show
isolated contributions from the distinct phonon mechanisms for the
$5.0$ meV QD; TA piezoelectric dominates at low fields. $(B)$.
Spin average values of the Zeeman sublevels after diagonalization
of $H$. Squares (circles) refer to ground (first excited) state.
$(C)$. Energy splitting of Zeeman sublevels after diagonalization
of $H$. Dashed-dotted lines show results without SO coupling.
Color scheme in panels $B$ and $C$ follow from panel $A$.}
\label{fi1}
\end{figure}

Panels $B$ and $C$ of Fig.\ \ref{fi1} respectively show the spin
expectation value $\langle \sigma _{z} \rangle$ of the Zeeman
sublevels and their energy splitting $\Delta E$. Smaller $E_{0}$
values show higher SO-induced spin mixing, so that $\langle \sigma
_{z}\rangle$ deviates from the pure $+1$ ($-1$) value for the
ground (first excited) state, and $\Delta E$ deviates from the
pure $g_{0}\mu _{B}B$ value of a QD without SO coupling. Notice at
the $3.5$ meV QD how a larger well (stronger field) decreases
(increases) the SO-induced spin mixing by reducing (enhancing) the
importance of the BIA (SIA) coupling. For comparison, the
experimental lower bound for the spin-flip time is 50 $\mu$s at
$B_{\parallel }=7.5$ T for a $1.1$ meV QD,\cite{hanson} while from
panel $A$ we find 5 $\mu$s. As discussed above, this time is
increased by one order of magnitude in a larger $z$-well, which
appears to be a better match for the experimental conditions. A 5
$\mu$s value has also been found from a perturbative
formulation,\cite{vitaly} and our results also agree with a
different model for the vertical confinement.\cite{rogerio} One
can then affirm that the perturbative approach\cite{nazarov} is
indeed enough for \textit{GaAs} in \textit{in-plane} fields at
\textit{any} QD sizes, and that the form of the $z$-confinement
has little effect on the rates.

In Fig.\ \ref{fi2} we do the same analysis for GaAs
QDs\cite{parga} in a perpendicular field, $B_{\perp}$. The most
striking feature is the appearance of minima in the rates shown in
panel $A$. The $E_{0}$-dependent minima ($12\leq B_{\perp }\leq
16$ T) are due to the \textit{vanishing} $\Delta E$, as shown in
panel $C$. At such values of $B_{\perp }$, a level crossing of
Zeeman sublevels results in a sudden \textit{spin-flip}, as
verified by the vertical lines of panel $B$ for the $0.7$ and
$3.5$ meV QDs. To understand the minima at low fields ($6\leq
B_{\perp }\leq 8$ T) we have to be mindful of the sine argument in
$F_{j}(z_{0})$, which may induce a minimum in the rate at a
particular value of the field, due to the interplay of energy and
length scales in the problem.\cite{ZeemanSO} This behavior does
not produce minima for in-plane field on GaAs QDs (Fig.\
\ref{fi1}). We find that those three low-field minima in Fig.\
\ref{fi2}$A$ occur at $B_{\perp }$ values where a \textit{magnetic
field-induced cancellation} of the SO influence is produced on the
respective $\Delta E$ values, as indicated by the arrows in panel
$C$: below (above) such $E_{0}$-dependent values of $B_{\perp }$
-- where $\Delta E$ recovers its Zeeman value of $g_{0}\mu _{B}B$
-- the SO coupling increases (decreases) the Zeeman splitting as
compared to the QD without SO.

\begin{figure}
\caption{Same as Fig.\ \ref{fi1}, but for GaAs QDs under
perpendicular magnetic field. Vertical lines in panel $B$ show the
field where a sublevel crossing occurs and the normal spin
character of the states is restored. Arrows in panel $C$ indicate
where suppression of SO coupling is induced by the magnetic field,
which is at the origin of the respective minima in panel $A$.}
\label{fi2}
\end{figure}

Notice that the two smallest confinements do not show the
low-field rate minima in Fig.\ \ref{fi2}$A$. This is due to the
lowest sublevels acquiring the same spin (between $3$ and $7$ T at
$E_{0}=0.7$ meV in panel $B$), so that the rate decreases
monotonically at that field-range. These features have strong
influence on QD effective $g$-factors.\cite{gfactor} Like in the
in-plane field problem, higher temperatures ($T=12$ K) enhance the
rates by one order of magnitude, and the dominating phonon
mechanism (symbols) is the TA piezoelectric coupling, but now at
\textit{any} field. To confirm the intricate interplay between all
QD energy scales, a larger well ($z_{0}=100$ \AA) or a stronger
field ($dV/dz=-38$ meV/\AA) \textit{removes} the rate minima, as
shown in panel $A$ for the $3.5$ meV QD. Our results agree with
available calculations\cite{cheng} at small fields ($B_{\perp
}\leq 1$ T), so that we can confirm that the perturbative approach
is \textit{not adequate} when dealing with SO effects in QDs in a
\textit{perpendicular} field, even if the material is GaAs,
\textit{if the vertical confinement is strong enough}.\cite{hdl}
From Figs.\ \ref{fi1} and \ref{fi2} the rates at small fields
($\lesssim 2$ T) have the same behavior under $B_{\parallel }$ and
$B_{\perp }$, namely, the smaller $E_{0}$ the larger the rate;
however, they are much larger under $B_{\perp }$. The rate maximum
(at 2.5 T for the 1.1 meV QD) is due to an anticrossing involving
the first and second QD excited states.\cite{gfactor}

\begin{figure}
\caption{InSb QDs\cite{parin} under in-plane magnetic field. The
confinements are such that the same QD size-range used for GaAs is
covered. In panel $A$, the $15.0$ ($20.0$) meV QD is chosen for
the study of different QD parameters (distinct phonon mechanisms);
dashed arrows show that both LA modes have the same number of rate
minima and happen at the same fields.} \label{fi3}
\end{figure}

Figure \ref{fi3} shows results for InSb QDs\cite{parin} in
in-plane fields. In contrast to GaAs (Fig.\ \ref{fi1}), minima in
the rates are visible under $B_{\parallel }$ for the highest
$E_{0}$ values. At small fields ($\lesssim 2$ T), a monotonic rate
drop is again observed as $E_{0}$ increases. As shown for the $20$
meV QD, the \textit{LA deformation potential} is the dominant
phonon mechanism for \textit{InSb} at \textit{any} field (and
$E_{0}$); this seems to be the case for any narrow-gap
material.\cite{gil} It is worth mentioning that, contrary to GaAs
in Fig.\ \ref{fi2}$A$ where a unique low-field minimum is present
for a given $E_{0}$, the number of rate minima in InSb increases
with $E_{0}$. Also, notice that both LA modes have the same number
of minima at the same $B_{\parallel}$ values (dashed arrows in
panel $A$), while a larger number of minima in the TA mode occurs
at different fields; this happens since $c_{LA}>c_{TA}$, so that
the sine argument in $F_{j}(z_{0})$ is smaller in the $LA$ mode.
Panel $B$ shows that Zeeman sublevels have normal spin character,
where ground (excited) state is predominantly spin-up (spin-down)
at low fields. Finite temperature ($T=12$ K) has no appreciable
effect on the rates, a reflection of the much larger energy scales
here. The energy scale interplay results at higher field
($dV/dz=-2$ meV/\AA) in slightly smaller (larger) $\Delta E$
(rates), and rate minimum shifts to higher fields. On the other
hand, a larger well ($z_{0}=100$ \AA) produces a higher $\Delta
E$, so that the spin-flip rate is suppressed by orders of
magnitude and its number of minima increases; notice in panel $B$
that the two lowest levels acquire the same spin at $\simeq 8$ T
for such $z_{0}$, so that the oscillatory rate is well defined
until that field. Previous calculations\cite{rogerio} in InSb did
not include the deformation potential, and yielded unrealistically
low rates.

For InSb QDs in perpendicular fields, we see from panel $B$ in
Fig.\ \ref{fi4} that all spin-flip rates in panel $A$ are
meaningful only for $B_{\perp }<8$ T, a field where both Zeeman
sublevels acquire the same spin. This happens earlier for small
$E_{0}$ as shown by the $3.0$ meV QD, whose rates are well defined
only for $B_{\perp }<2$ T; the vertical line around $1$ T in panel
$B$ for this $E_{0}$ indicates a spin-flip, accompanied by the
vanishing $\Delta E$ in panel $C$, the rate minimum in panel $A$,
and a sign change of the QD $g$-factor.\cite{gfactor} Larger
values of $E_{0}$ present normal spin behavior. Spin-flip rates in
panel $A$ do {\textit not} show the monotonic behavior seen for
small fields in Fig.\ \ref{fi3}. Like in the $B_{\parallel}$ case,
a finite temperature ($T=12$ K) does not show visible influence on
the rates under $B_{\perp }$, and the deformation potential
dominates at any field. A higher electric field ($dV/dz=-2$
meV/\AA ) slightly increases the rate; a larger well ($z_{0}=100$
\AA ) \textit{introduces} an oscillatory rate orders of magnitude
smaller. We emphasize that the perturbative approach finds
\textit{no use} in InSb QDs because of the inherent higher SO
coupling. It is worth mention that oscillatory rates have also
been found in GaAs QDs under $B_{\perp }$ by considering
\textit{momentum} relaxation from an excited orbital level to the
ground state,\cite{bock} as well as in coupled GaAs
QDs.\cite{bert}

\begin{figure}
\caption{Same as Fig.\ \ref{fi3}, but for InSb QDs under
perpendicular magnetic field. The only minimum of the $3.0$ meV QD
around $1$ T in panel $A$ indicates a sublevel crossing, as shown
by the vertical line in panel $B$ and the vanishing of $\Delta E$
in panel $C$. The color scheme is the same in all panels and
figures.} \label{fi4}
\end{figure}

Even though both QD materials show oscillatory spin-flip rates --
with $B_{\perp}$ for GaAs and with both $B_{\parallel}$ and
$B_{\perp}$ for InSb -- their origin is slightly different. Minima
come from the nature of the $z$-confinement, and the field where
they occur depend on the lateral size $l_{0}$. The SO coupling
mixes spins and alters splitting of sublevels in distinct ways
according to field-direction and QD material, so that spin
relaxation can be induced by piezoelectric (wide-gap) and
deformation (narrow-gap) phonons. Our calculations reveal the rich
interplay between all relevant energy scales in QDs. The external
field opens channels (at the rate minima) where long spin
relaxation times ($\simeq 1$ s) may be reached, so that the spin
coherence required for quantum computing could be improved.

We thank support from NSF-IMC grant 0336431, CMSS at OU, and the
$21^{st}$ Century Indiana Fund.


\begin{thebibliography}{99}
\bibitem{parga} In GaAs: $m=0.067$ m$_{0}$, $g_{0}=-0.44$, $\kappa
=12.4$, $\alpha=4.4$ \AA $^{2}$, $\gamma =26$ eV\AA $^{3}$, $\Xi
_{0}=7$ eV, $eh_{14}=0.140$ eV/\AA , $N_{0}=5.32\times 10^{-27}$
Kg/\AA $^{3}$, $c_{LA}=4.73\times 10^{13}$ \AA /s, and
$c_{TA}=3.35\times 10^{13}$ \AA /s; also, $z_{0}=40$ \AA ,
$dV/dz=-0.5$ meV/\AA , and $T=0$ K if no other numbers are
specified.
\bibitem{hdl} $\left\langle k_{z}^{2}\right\rangle =(\pi /z_{0})^{2}$
for an infinite well. If a Gaussian well is considered, one has
$\left\langle k_{z}^{2}\right\rangle =1/z_{0}^{2}$, which yields a
BIA term a factor of $10$ smaller than used in this work.
\bibitem{parin} In InSb: $m=0.014$ m$_{0}$, $g_{0}=-51$, $\kappa
=16.5$, $\alpha =500$\AA $^{2}$, $\gamma =160$ eV\AA $^{3}$, $\Xi
_{0}=7$ eV, $eh_{14}=0.061$ eV/%
\AA , $N_{0}=5.77\times 10^{-27}$ Kg/\AA $^{3}$,
$c_{LA}=3.40\times 10^{13}$\AA /s, and $c_{TA}=2.29\times 10^{13}$
\AA /s; also, $z_{0}=40$ \AA , $dV/dz=-0.5$ meV/\AA , and $T=0$ K
if no other numbers are specified.
\end{thebibliography}
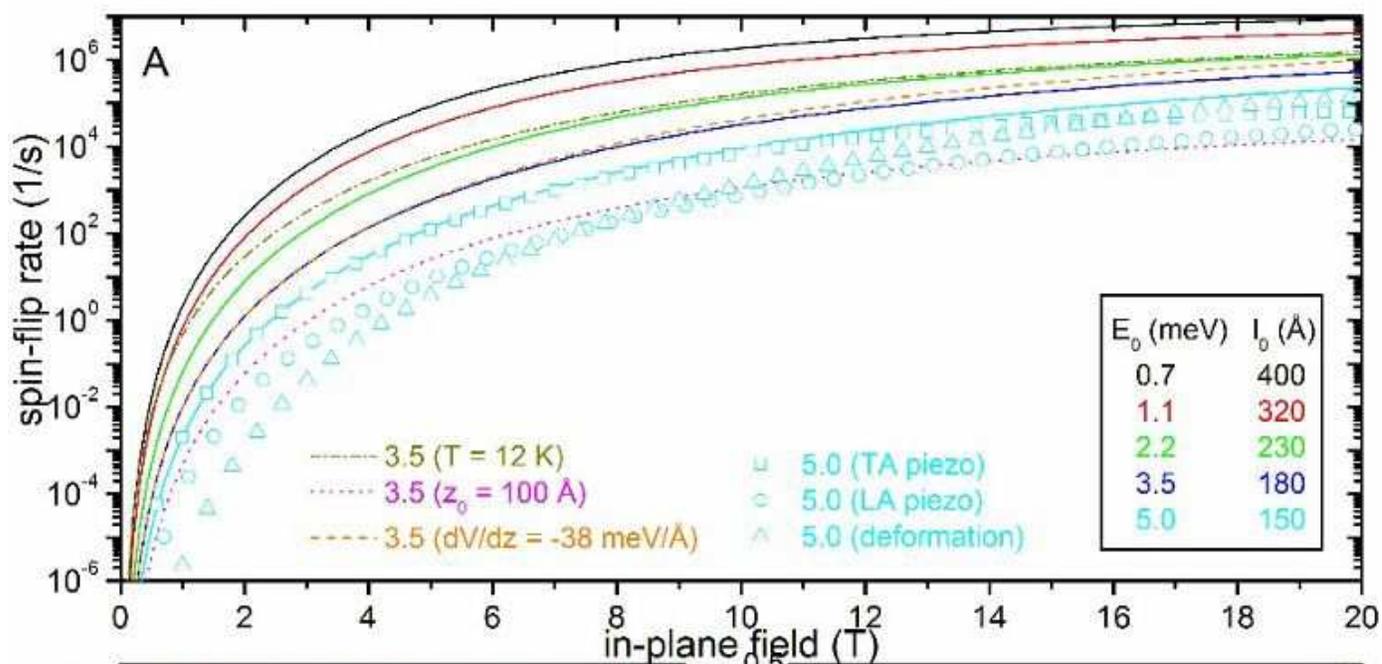
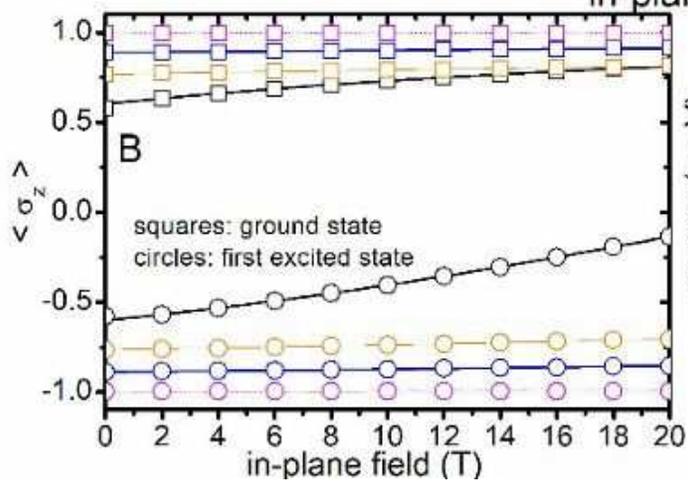
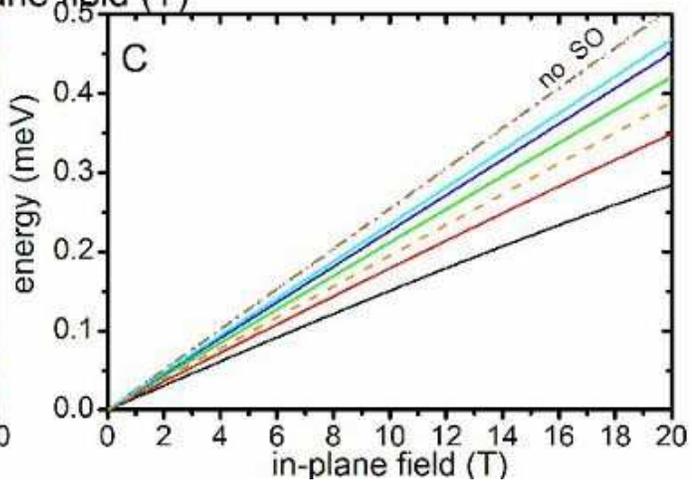
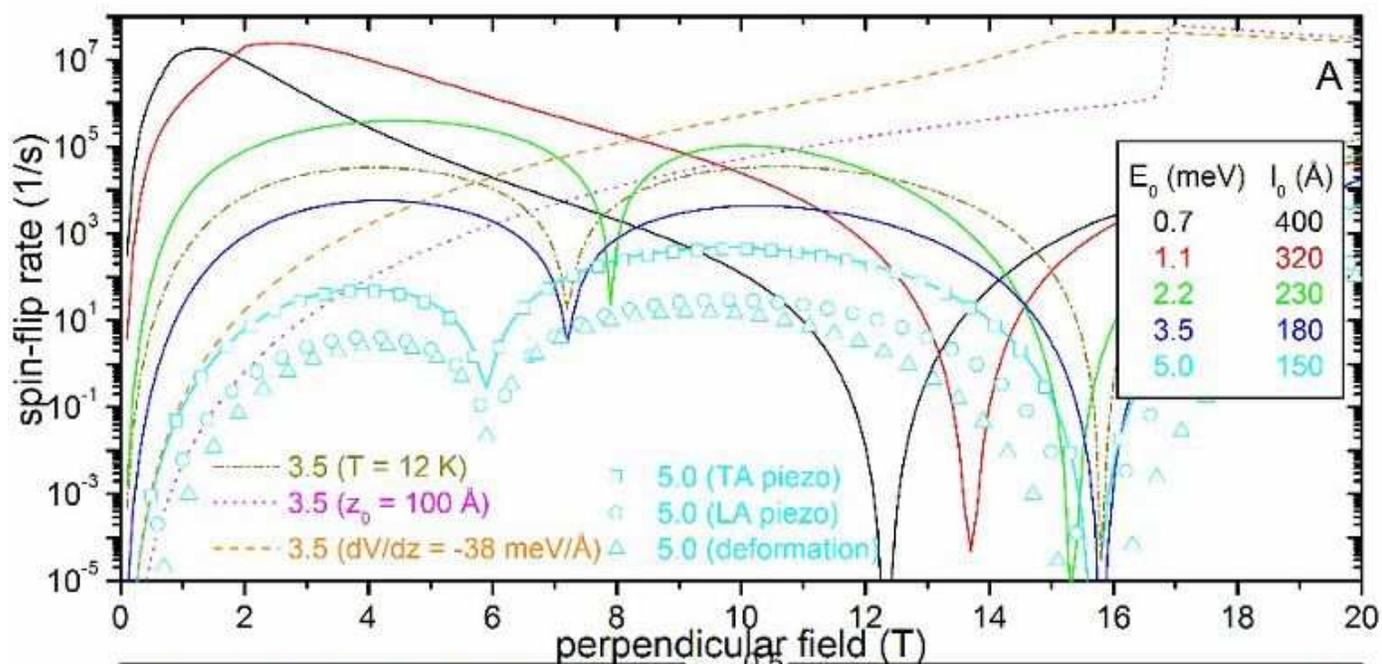
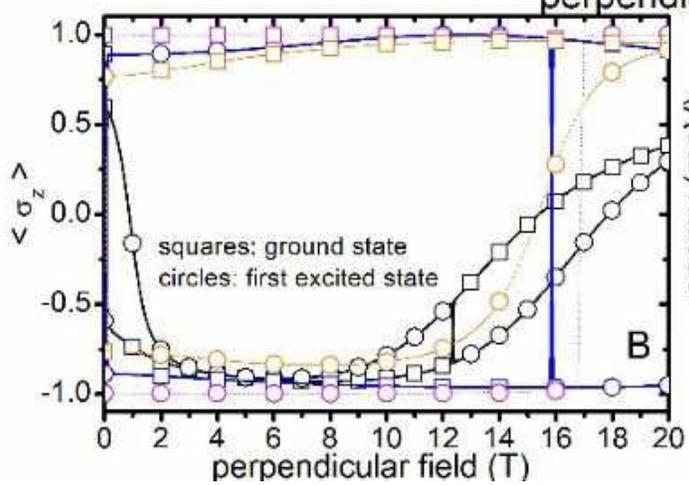
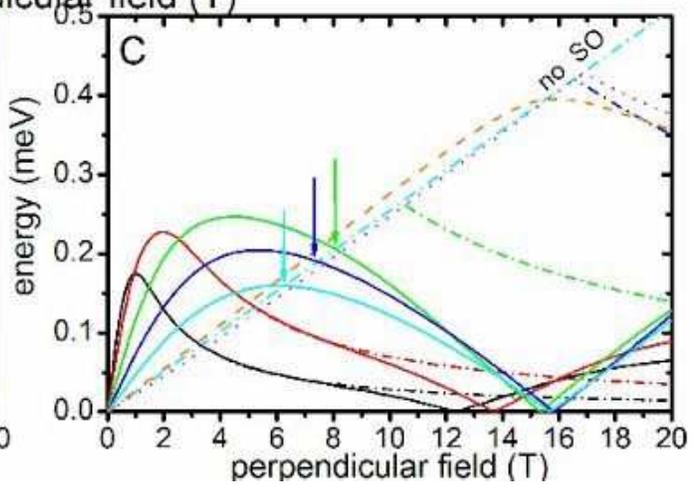
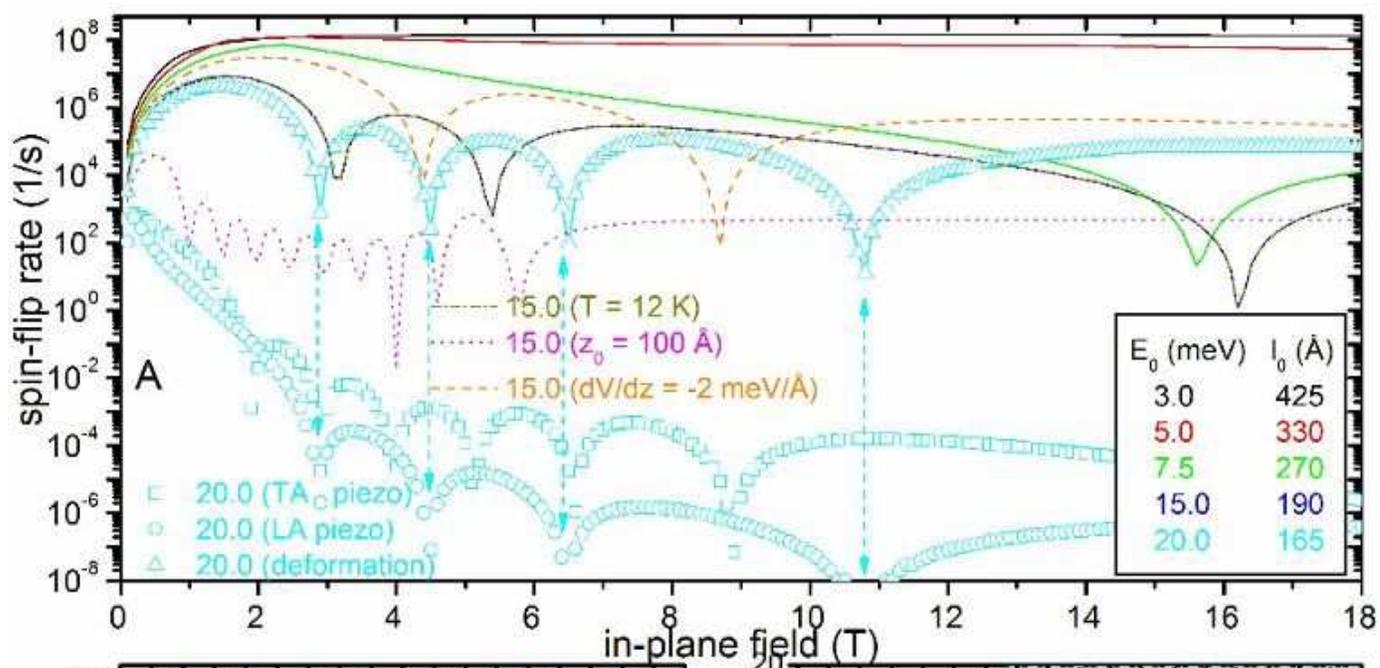
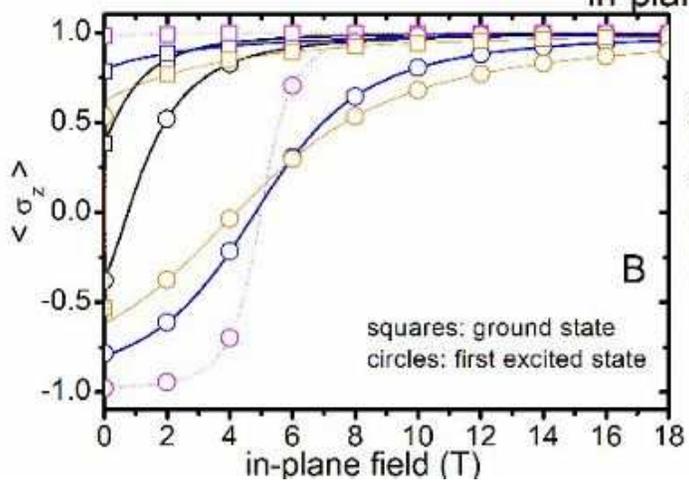
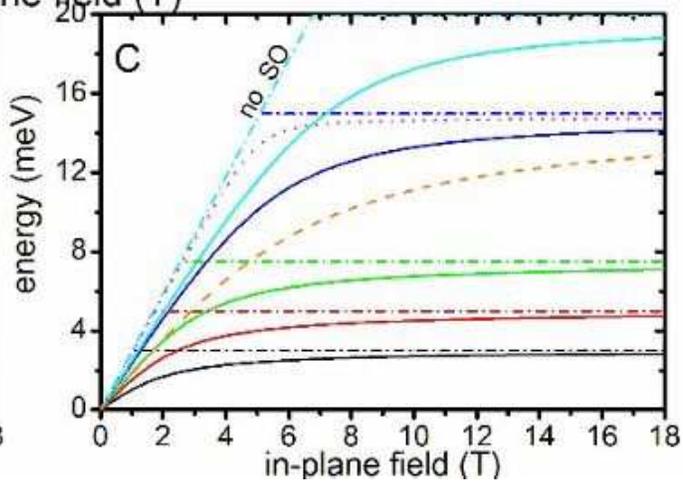
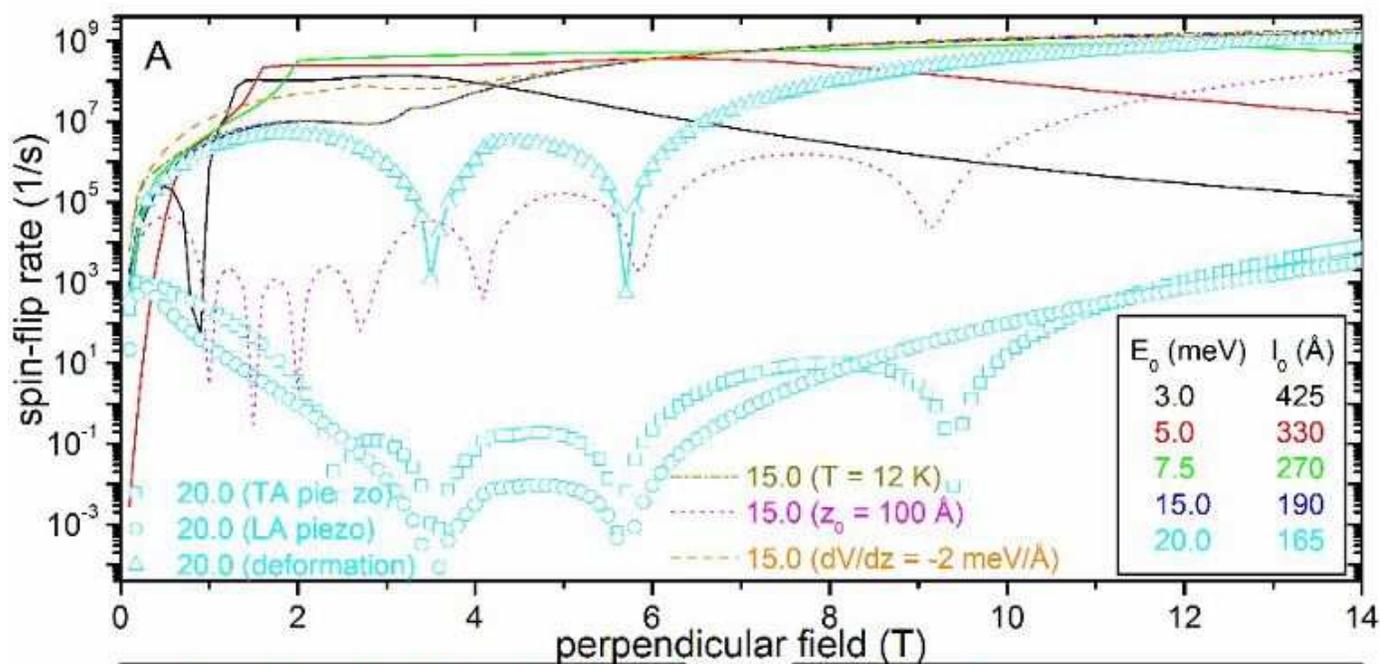
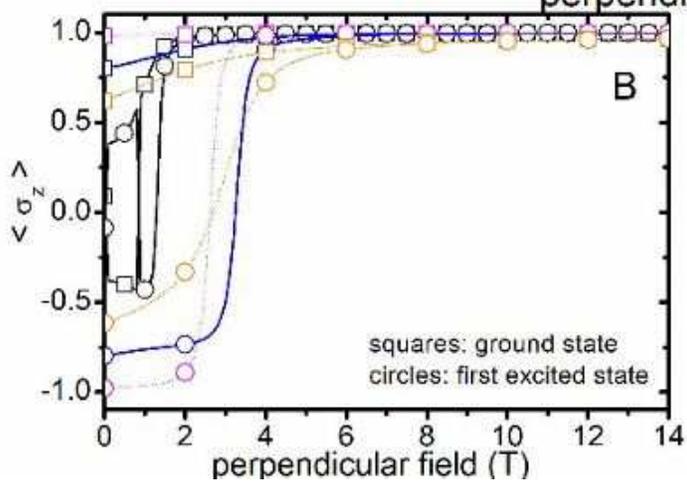
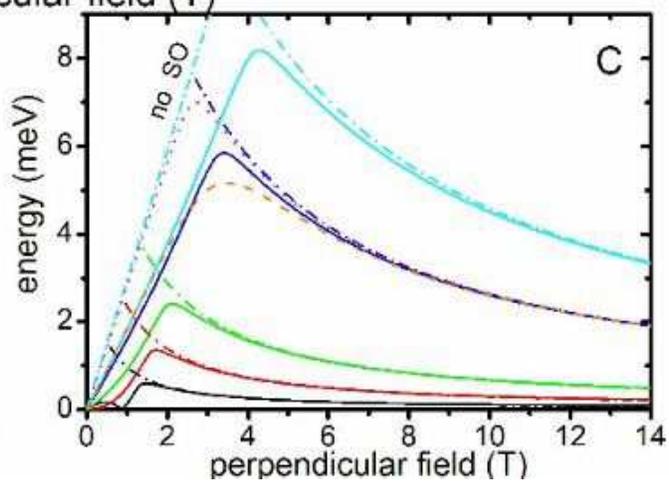
\end{document}